\documentclass{article}
\usepackage[pdftex]{graphicx}        
\usepackage{natbib}         
\pdfoutput=1
\newcommand{\etal}{{\it et al.}}

\newcommand{\aap}{    {\it Astron. Astrophys. }}

\newcommand{\apj}{    {\it Astrophys. J.}}

\newcommand{\mnras}{  {\it Mon. Not. Roy. Astron. Soc.}}
\newcommand{\nat}{    {\it Nature}}

\newcommand{\solphys}{{\it Solar Phys.}}

\title{Seismic Emissions from a Highly Impulsive M6.7 Solar Flare}

\author{J.C. Mart\'inez-Oliveros,
        H.   Moradi,
        A-C. Donea\\
\vspace{2mm} \\
{\it Centre of Stellar and Planetary Astrophysics, Monash University,} \\
 {\it                 Victoria 3800, Australia}\\
 {\it                 email:Juan.Oliveros@sci.monash.edu.au}
}

\begin{document}

\maketitle
             
\begin{abstract}

On 10 March 2001 the active region NOAA 9368 produced an unusually impulsive solar flare in close proximity to the solar limb. This flare has previously been studied in great detail, with 
observations classifying it as a type 1 white-light flare with a very hard spectrum in hard 
X-rays. The flare was also associated with a type II radio burst and coronal mass ejection. 
The flare emission characteristics appeared to closely correspond with previous instances of seismic emission from acoustically active flares. Using standard local helioseismic methods, we identified the seismic signatures produced by the flare that, to date, is the least energetic (in soft X-rays) of the flares known to have generated a detectable acoustic transient. Holographic analysis of the flare shows a compact acoustic source strongly correlated with the impulsive hard X-ray, visible continuum, and radio emission. Time-distance diagrams of the seismic waves emanating from the flare region also show faint signatures, mainly in the eastern sector of the active region. The strong spatial coincidence between the seismic source and the impulsive visible continuum emission reinforces the theory that a substantial component of the seismic emission seen is a result of sudden heating of the low photosphere associated with the observed visible continuum emission. Furthermore, the low-altitude magnetic loop structure inferred from potential--field extrapolations in the flaring region suggests that there is a significant inverse correlation between the seismicity of a flare and the height of the magnetic loops that conduct the particle beams from the corona.

\end{abstract}

\section{Introduction}
     \label{S-Introduction} 

Recent developments in the study of flare acoustic emissions \citep{donea05,donea06,moradi07,Martinez-Oliveros2006} have bolstered the view that seismic emission from flares offers major new insights both into flare physics and helioseismology, ranging from a greatly improved understanding of flare dynamics and kinematics, to an understanding how seismic emission is generated differently by turbulence in magnetic subphotospheres from inside the quite Sun. 

Sunquakes emanate from compact sources that take only a small fraction of the energy released by flares. The surface manifestation of these sources appears as circular (or near-circular) waves propagating outward from the solar surface, approximately 20\,--\,60 minutes after the impulsive phase of the flare. \citet{donea05} considered the possibility that relatively weak flares might be able to produce sunquakes and that acoustically active flares may indeed be more common than previously thought. This was confirmed soon after by \citet{betal2006} following comprehensive helioseismic observations of flares using helioseismic holography and the data from Michelson Doppler Imager (MDI) onboard the Solar and Heliospheric Observatory (SOHO). 

While the majority of the flare acoustic transients discovered to date have been released by the more energetic X-class flares, recently however, a number of strong acoustic emissions from M-class flares have been discovered. \citet{donea06} analyzed the helioseismic properties of the strong seismic transient produced by the M9.5 class flare of 9 September 2001, and very recently,  \citet{Martinez-Oliveros2006} performed a comprehensive electromagnetic and acoustic analysis of the M7.4-class flare of 14 August 2004 -- the smallest flare known to have produced a detectable acoustic transient prior to the discovery of the sunquake reported in this paper. Their findings, along with those of \citet{donea05} and \citet{moradi07}, have identified a number of distinct observational characteristics that distinguish acoustically active flares from others: 

\begin{enumerate}
\item The sites of seismic emission generally coincide spatially with impulsive hard X-ray (HXR) and microwave (MW) emissions, suggesting a relation to thick-target heating of the chromosphere by energetic particles.
\item The sites of seismic emission similarly coincide spatially with impulsive continuum emission, suggesting acoustic emission associated with extra heating and ionization of the low photosphere. 
\item The seismicity of the active region appears to be closely related to the heights of the coronal magnetic loops that conduct high-energy particles.  
\end{enumerate}

In this paper, we will examine the last of the above characteristics -- introduced by \citet{Martinez-Oliveros2006} -- in greater detail, with further evidence from the flare of 10 March 2001. The evidence we will provide reinforces the theory that shorter coronal loops are more likely to be conducive to a more rapid injection of trapped, high-energy electrons into the chromosphere at their footpoints. This enhances the magnitude and suddenness of the chromospheric heating that gives a rise to the intense visible continuum emission seen in all acoustically active flares. This mechanism appears to be a prospective source of the energy required to drive a powerful acoustic transient into the solar interior.

\section{The Helioseismic Signatures}

The helioseismic analysis relies on raw data from SOHO/MDI. The data consist of full-disc Doppler, magnetogram, and continuum images in the photospheric line Ni\,{\sc i}~6768~\AA, obtained at a cadence of one minute. The Dopplergrams were corrected for small effects due to reduced oscillatory amplitudes in magnetic regions, following the method outlined in \citet{raj2006}. We utilize two different, but in our opinion, complementary helioseismic techniques to analyse the seismicity of the acoustic emission produced by the flare. 

\begin{figure}[ht]
\begin{tabular}{cc}
\includegraphics[width=0.45\columnwidth]{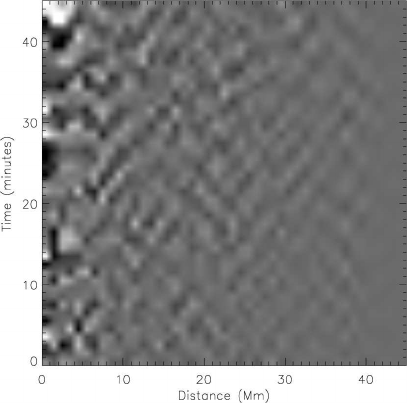}&
\includegraphics[width=0.45\columnwidth]{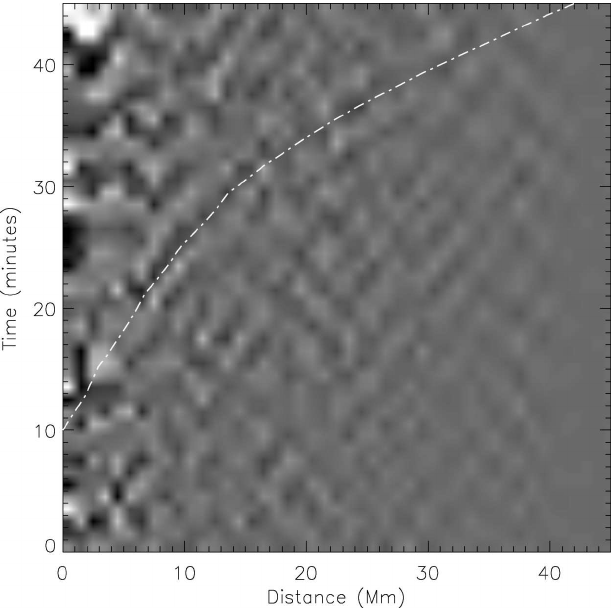}
\end{tabular}
\caption{Time\,--\,distance plot of the amplitude of the surface ridge averaged over curves of constant radius in the azimuth range $+135^o$ to $+225^o$ are rendered in gray in both frames. The white curve superimposed in the right frame represents the wave time travel for a standard model of the solar interior. The time represented as 0 along the vertical axis of the plot is 04:07~UT.}
\label{td_plot}
\end{figure}

The first method employed was the time--distance technique described by \citet{kz1998}. We generate the time--distance plot over a selected range of azimuths from the primary HXR and magnetic--transient sources, in this case $+135^o$ to $+225^o$, in order to gauge the expanding signal from this region and compare this signal with a curve that represents the theoretical group travel time. The resulting signature, manifested as a ``ridge'' in the time-distance diagram, was significant, but as was expected, appeared to be quite weak (see Figure~\ref{td_plot}). This is more than likely a consequence of the relatively small energy released by the flare (class M6.7 in X-rays) that produced the sunquake. 

The theoretical curve appears to match the observed ridge with a delay of approximately 5 minutes from the time of the flare maximum. A temporal delay of such nature was contemplated in \citet{zharkova2007}, in our case being of slightly longer duration. According to \citet{zharkova2007}, this delay is due to the time required for the electrons to move along the magnetic field lines and hit the upper photosphere or chromosphere. The velocity and acceleration of the expanding wave packet was also computed. The velocity of the wave-front was calculated to be between 5 and 9~Mm was $\approx 13~\mathrm{ km\,s^{-1}}$ and between 29 and 33~Mm, $\approx 66.67~\mathrm{ km\,s^{-1}}$. The mean acceleration of the wave-front was also estimated to be $\approx 3.35~\mathrm{ km\,s^{-2}}$. 

The second method employed in our analysis was computational seismic holography, to image the acoustic source of the sunquake. This method has been used extensively in the analysis of acoustically active flares, with great success in identifying numerous seismic sources from solar flares \citep{donea99,donea05,donea06,moradi07,Martinez-Oliveros2006}. Helioseismic holography can be described as  essentially the phase-coherent reconstruction of acoustic waves observed at the solar surface into the solar interior to render stigmatic images of subsurface sources that have given rise to the surface disturbance. In general, the acoustic reconstruction can be done either forward or backward in time. When it is backward in time, we call the extrapolated field the``acoustic egression''. In the case of subjacent--vantage holography this represents waves emanating from surface focus downward into the solar interior that have subsequently refracted back to the surface in an annular pupil surrounding the source.  For the sake of brevity, we direct the reader to \citet{lb2000} for a more in-depth discussion of holographic techniques.

 \begin{figure}[!h]
 \begin{center}
 \includegraphics[width=0.95\columnwidth]{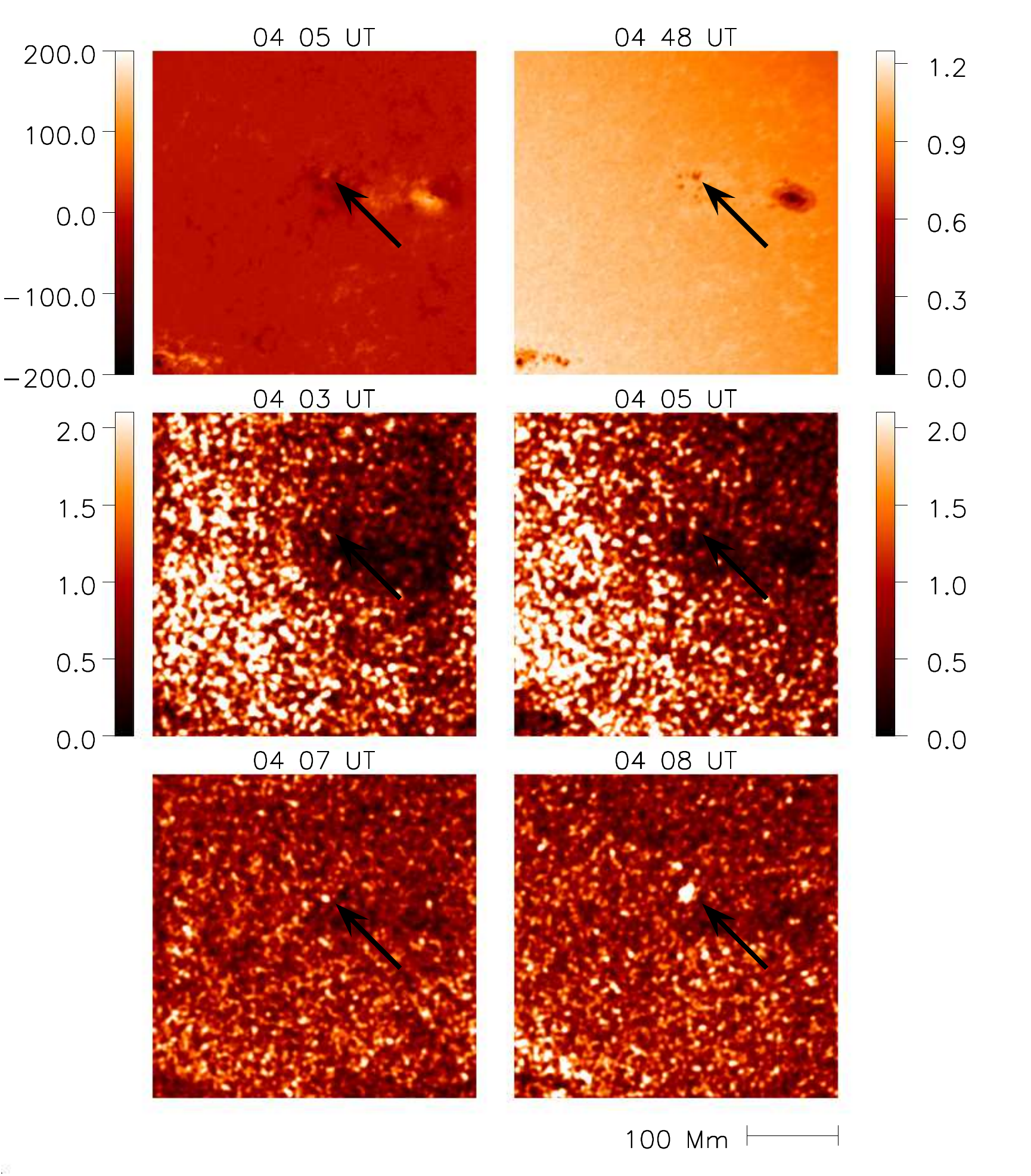}
 \caption{Egression and acoustic power snapshots of AR~9348 on 10 March 2001 integrated over 2.0\,--\,4.0~mHz and 5.0\,--\,7.0~mHz frequency bands and taken at the maximum of the correspondence frequency. Top frames show MDI magnetogram of the active region (right) at 04:05~UT and a visible continuum image at 04:08~UT(left). Second row shows egression power at 3~mHz (left) and 6~mHz at the respective maxima. The bottom row show acoustic power. Times are indicated above respective panels, with arrows inserted to indicate the location of the seismic source.}
 \label{snap_egre}
 \end{center}
 \end{figure} 

To assess the seismic emission from the flare, we computed both the acoustic and egression power over the neighbourhood of the active region at one-minute intervals, mapping them for each minute of observation. It is important to distinguish between the ``egression'' power -- wherein each pixel is a coherent representation of acoustic waves that have emanated downward from the focus, deep beneath the solar surface, and re-emerged into a surrounding annular pupil, and the local ``acoustic'' power  -- wherein each pixel represents local surface motion as viewed from directly above the photosphere. 

The resulting acoustic and egression power movies and ``snapshots'' (acoustic/ egression power sampled over the solar surface at any definite time) are computed over 2 mHz bands, centred at 3 and 6 mHz. The higher frequency band has a number of advantages because it avoids the much greater ambient noise of the quiet Sun that predominates the 2\,--\,4 mHz frequency band and due to a shorter wavelength, it also provides us with the images that have a finer diffraction limit.

Acoustic and egression power snapshots at the maximum of the flare are shown in Figure~\ref{snap_egre}. In these computations, the pupil was an annulus of radial range 15\,--\,45 Mm centred on the focus. To improve the statistics, the original egression power snapshots are smeared by convolution with a Gaussian with a 1/$\mathrm e$-half-width of 3 Mm. The egression power images and the continuum image are also normalized to unity at respective mean quiet-Sun values.  The acoustic signature of the flare -- consisting of a bright compact source -- is clearly visible at 6 mHz in the both acoustic and egression power snapshots at 04:05~UT (indicated by the arrows in  Figure~\ref{snap_egre}). At 3 mHz the egression and local acoustic power snapshots show a less conspicuous signature than at 6 mHz due to a much greater background acoustic power at 3 mHz.

\begin{figure}[!h]
\begin{center}
\includegraphics[width=0.95\columnwidth]{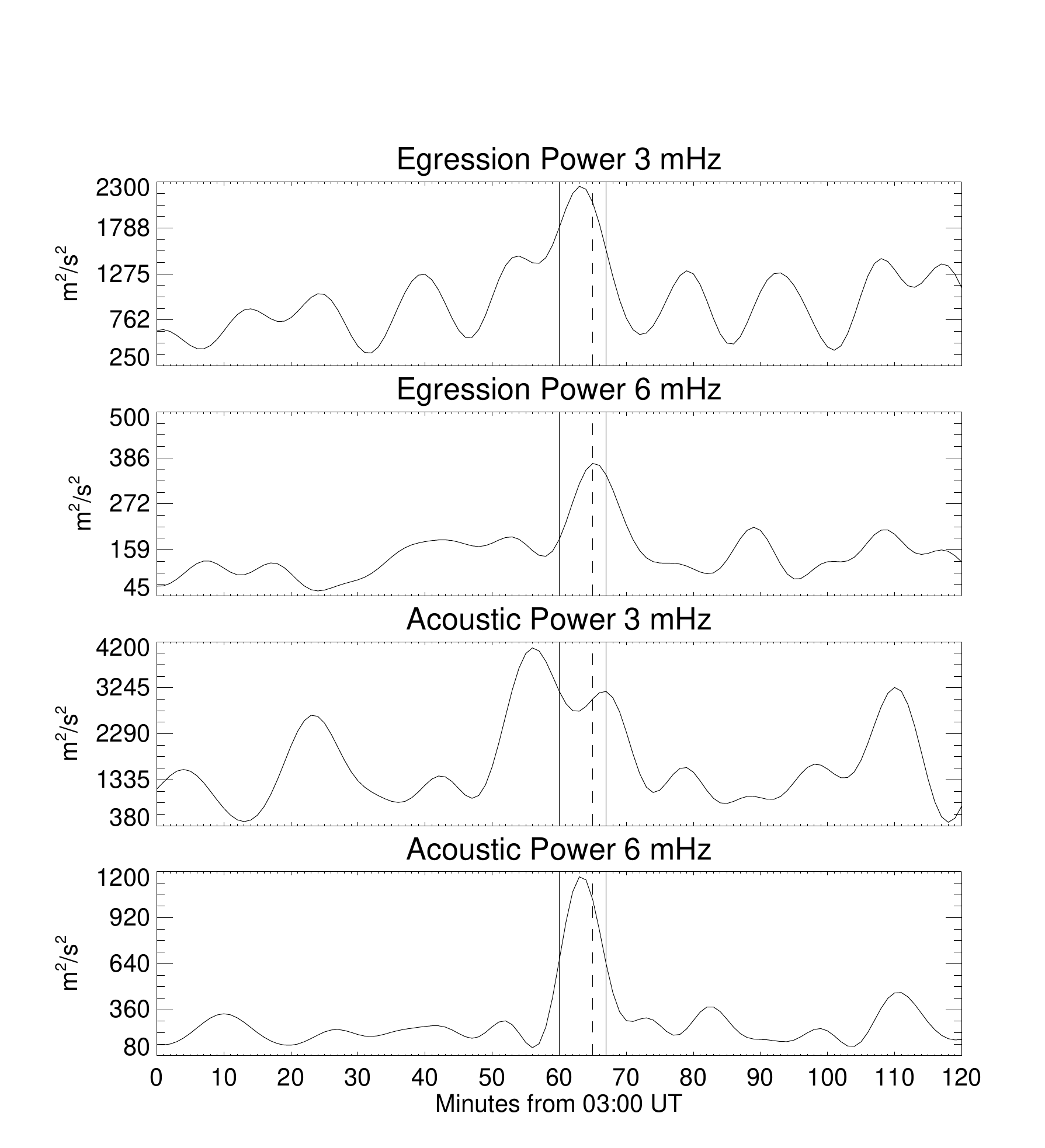}
\caption{The 3- and 6-mHz egression and acoustic power time series, integrated over the neighbourhood of the egression power signatures. The vertical lines represent the beginning (04:00~UT), maximum (04:05~UT), and end (04:07~UT) times of the GOES X-ray flare.}
\label{egression}
\end{center}
\end{figure}

The temporal profiles of the seismic source, seen in the acoustic/egression time-series in Figure~\ref{egression} correspond closely with other compact manifestations of the flare including significant white-light (WL) emission with a sudden, impulsive onset as discussed by \citet{lietal2005} and \citet{uddinetal2004}. 
The spatial and temporal features of the seismic source observed also coincides closely with the HXR signature reported by \citet{lietal2005}, indicating that high-energy particles accelerated above the chromosphere contribute to the generation of the seismic source. We will discuss their observations in more detail in the next section. 


\section{Multi-Wavelength Analysis}
\label{multi}

The multi-wavelength properties of the extremely impulsive WLF of 10 March 2001, have previously been studied in detail by a number of authors \citep{liuetal2001, ding03, uddinetal2004,lietal2005};  all emphasizing the impulsiveness of the flare and the strong spatial and temporal coincidence of the hard HXR emission with the enhanced continuum emission. 

\begin{figure}[!h]
\begin{center}
\includegraphics[width=0.8\columnwidth]{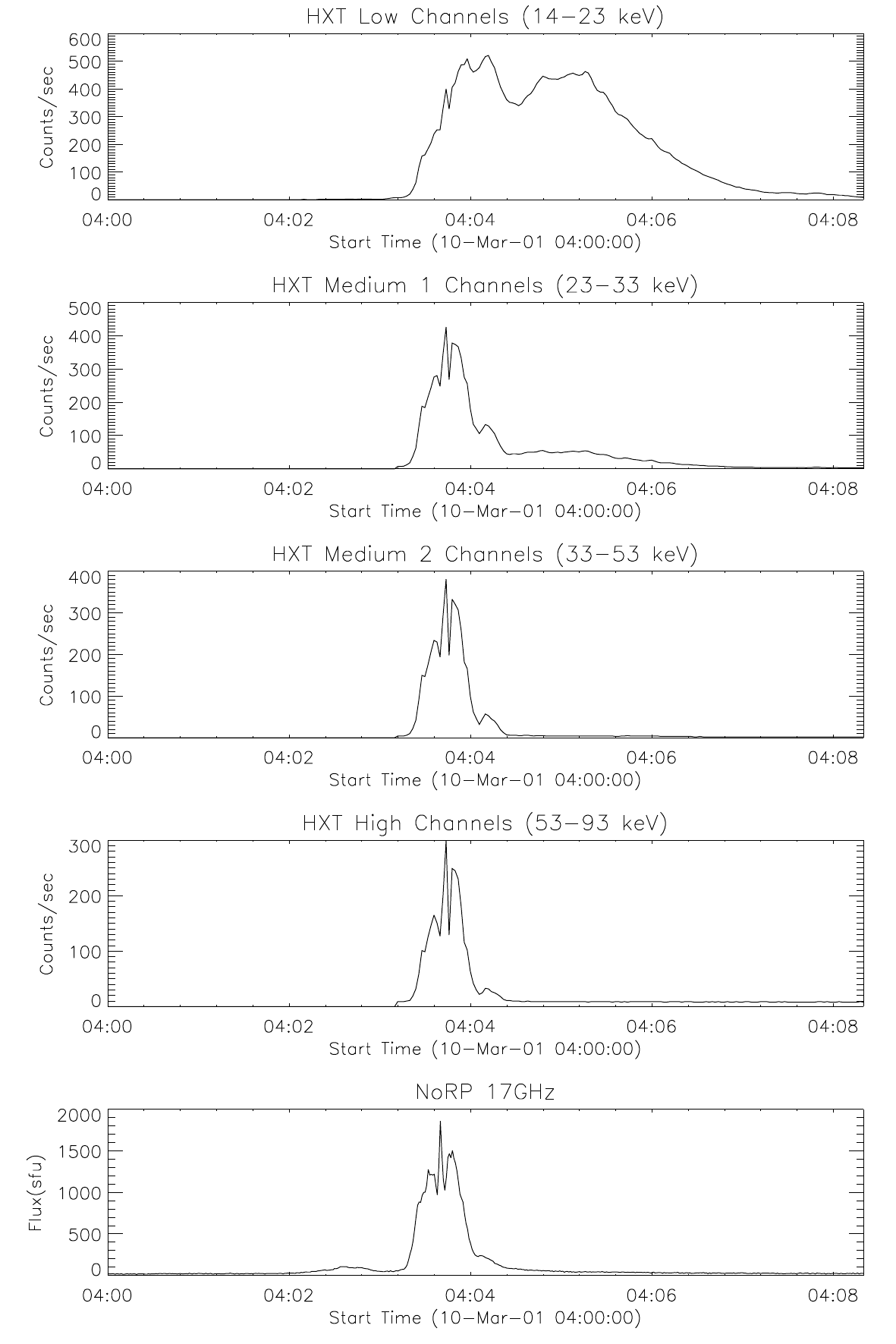}
\caption{HXR and MW time profiles. The HXR fluxes were taken by {\it Yohkoh} at the L(14\,--\,23~keV), M1(23\,--\,33~keV), M2(33\,--\,53~keV), and H(53\,--\,93~keV) channels. The NoRP flux plotted correspond to the 17~GHz channel.}
\label{flujos}
\end{center}
\end{figure}

The observations of \citet{uddinetal2004} showed that the flare embodied a very hard spectrum in HXR, a type II radio burst, and a coronal mass ejection. GOES SXR observations classified it as a M6.7 class, beginning at 04:00~UT, reaching its maximum at 04:05~UT, and ending at 04:07~UT.  A very important characteristic of the flare of 10 March 2001 is its duration, which was approximately seven minutes, indicating that the physical processes associated with the flare also had a very short duration. \citet{uddinetal2004} made a detailed study of this flare at different wavelengths and determined that all three main phases of the flare could be observed clearly in different temporal profiles in HXR at different energy bands (Figure~\ref{flujos}). 
The precursor phase was observed to occur at 04:03~UT with a duration of 15 seconds, the impulsive phase between 05:03:15 and 04:03:40~UT, and the gradual phase after 04:03:40~UT. Also, they calculated the column emission measure, the spectral index of the flare signal and the temporal variation of the temperature. They found that the emission has a non-thermal component before 04:04~UT and thermal component after 04:05~UT. From the observed profiles, they concluded that a very fast acceleration of the electrons occurs during the impulsive phase.

\citet{uddinetal2004} also emphasized the spatial and temporal correlation of the HXR source and the continuum emission. They also commented on the uncommon change of magnetic flux they detected, concluding that it indicates that the WLF was triggered by a new emerging flux that induces a flux cancellation. As a result, they conclude that magnetic reconnection occurred in the upper atmosphere of the sunspot region; thereby high-energy electrons precipitate along magnetic field lines and deposit energy at the sunspot region, which produce the HXR and continuum enhancement. 

\begin{figure}[!h]
\begin{center}
\includegraphics[width=0.80\columnwidth]{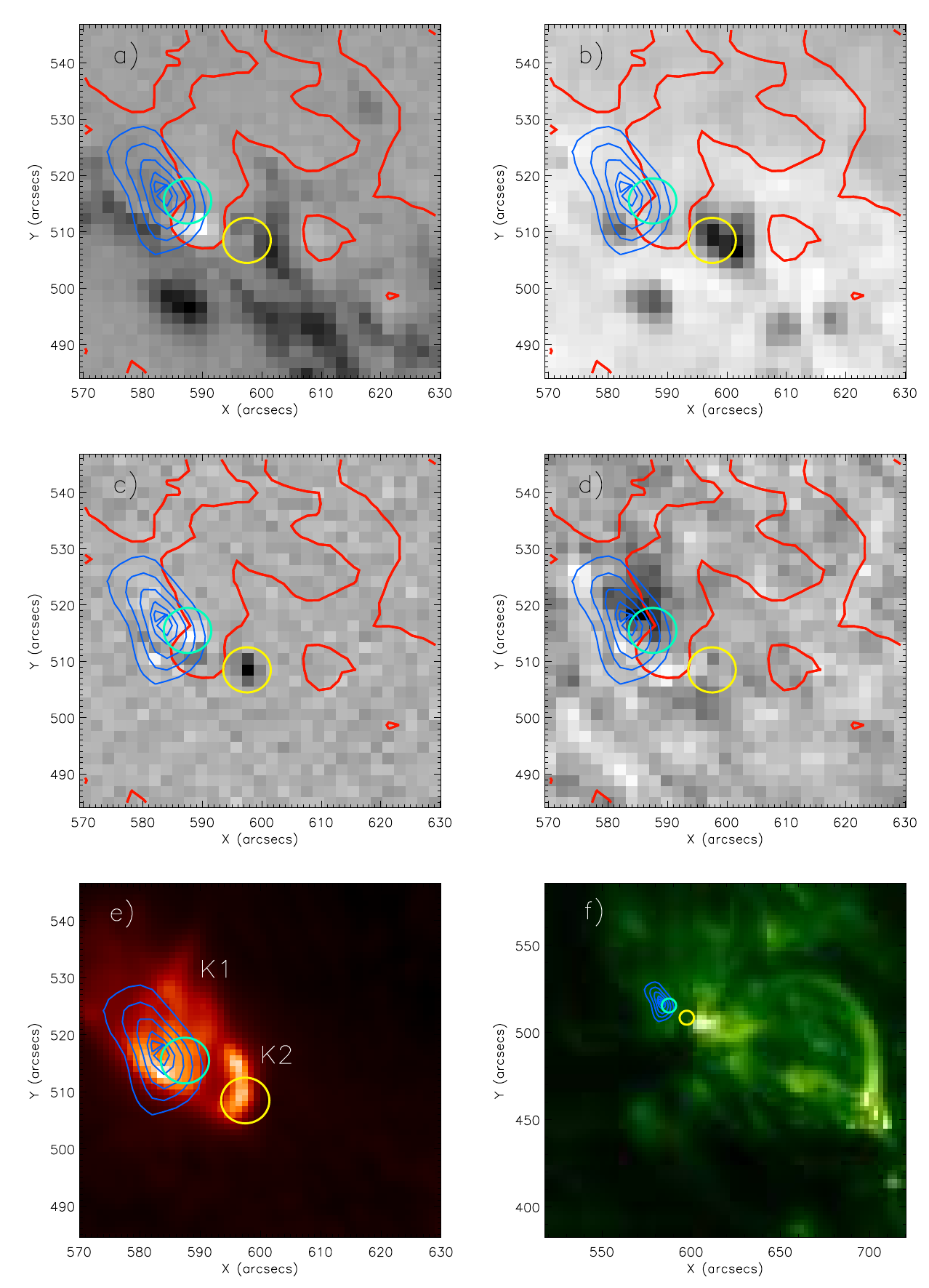}
\caption{HXR contours of the flare at 04:03:38~UT overlaid over a) MDI intensity continuum, b) MDI magnetogram, c) MDI magnetogram difference at the flare maximum, d) Doppler difference at the maximum, e) $\mathrm H\alpha$ and f) SOHO/EIT at 171~\AA. The background images all correspond to the same time (04:04:01.61~UT). The HXR contour levels are 20, 40, 60, 80, and 90\% of the maximum emission in the M2 (22-53~keV) channel. The MDI magnetogram neutral line (red line) is overlaid  in the frames a), b), c) and d). The blue and yellow circles in all frames represent the relative position of the main magnetic transients. The seismic source coincides spatially with the blue circle, where there is also the HXR emission.}
\label{mosaic2}
\end{center}
\end{figure}

The importance of this particular type of spatial and temporal correlation between the different types of multi-wavelength signatures described above, in the presence of a seismic source, was first identified and discussed in depth by \citet{Martinez-Oliveros2006}. They identified a significant temporal correlation between the fluxes at different frequencies and energy bands (for the M7.4 class flare of 14 August 2004) which were seen to be directly related to two electron populations - one trapped in the magnetic field, and another precipitating into the chromosphere. The highly impulsive character of this flare indicates that the trapped population of electrons in the magnetic field was injected into the chromosphere very fast. The electrons had no time to thermalize in the coronal loop, but were evacuated by rapid precipitation, therefore they did not produce a significant emission in MW. Indeed, this type of emission is absent in the MW profile reported by \citep{uddinetal2004}. 
The radio emission does not show a long exponential decay, implying that high-energy electrons that are generally trapped for a significant amount of time in long coronal loops that extend to great heights, are evacuated by rapid precipitation in short, low-lying loops.

\citet{lietal2005} also observed the WL properties of the 10 March 2001 flare, detecting an infrared continuum enhancement of 4\,--\,6\% compared to pre-flare values. The study of the continuum images shows that the WL source is located over the magnetic neutral line and that the source is most likely composed of the two footpoints of the magnetic loop, which are too close together to be resolved by the RHESSI HXR observations. They also detected a HXR source near the sunspot. The authors also concluded from their observations that the temporal and spatial coincidence of the HXR emission with the continuum emission indicates that electron precipitation may have been the main energy source of the chromospheric heating, producing the excess continuum emission. Furthermore, they suggest that the electron-beam bombardment, coupled with radiative back-warming effects, plays the main role in the heating of the sunspot atmosphere. This is significant because all instances of seismic emissions to date have exhibited very similar WLF characteristics - characterized in particular by the sudden appearance of the WL signature during the impulsive phase of the acoustically active flare. 

\begin{figure}[!h]
\begin{center}
\includegraphics[width=0.95\columnwidth]{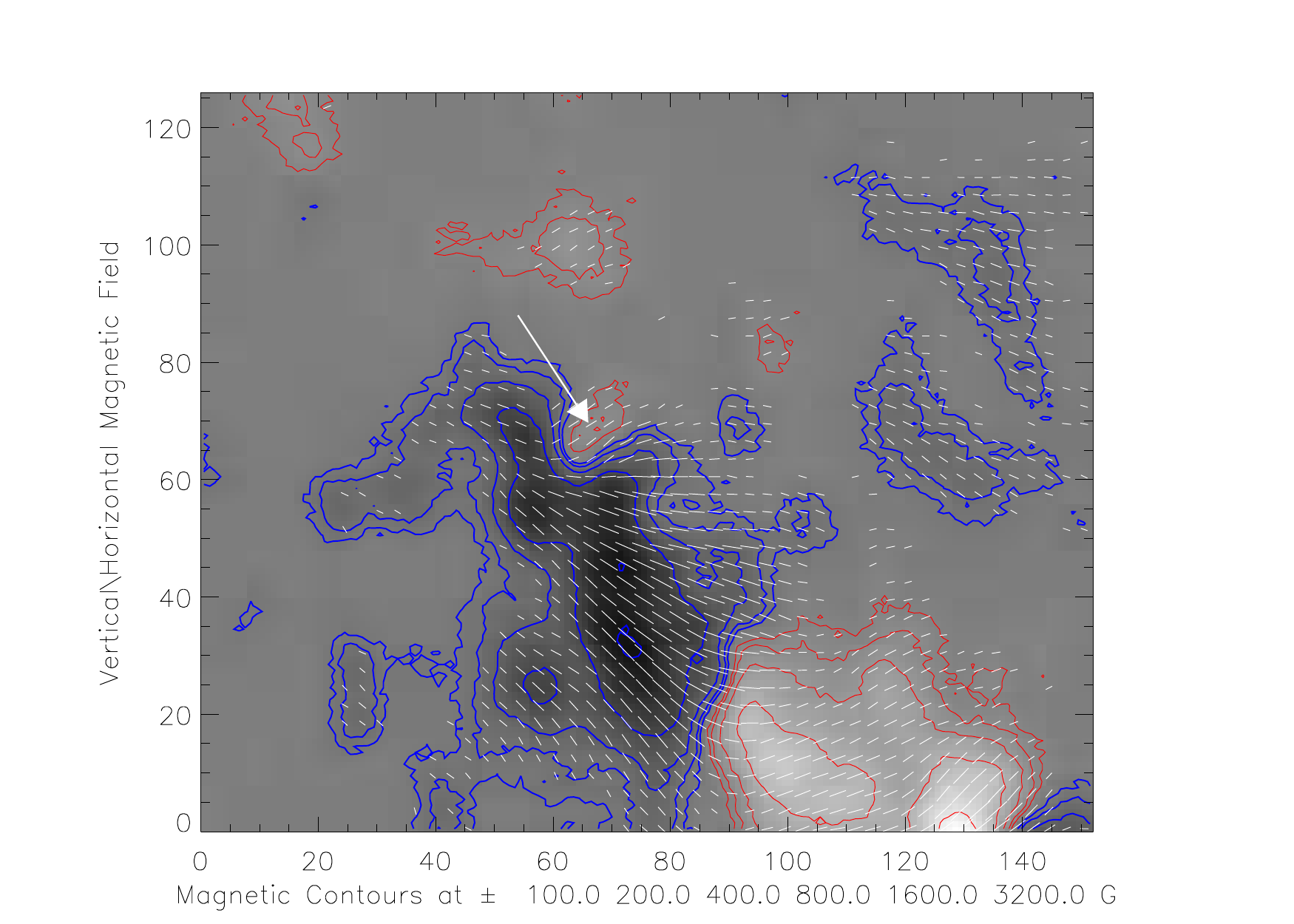}
\caption{Vector magnetogram of the active region taken by the Mitaka Observatory (NAOJ) at 00:10:16~UT. The sunquake region is pointed with an arrow.}
\label{vm}
\end{center}
\end{figure}

The images in Figure~\ref{mosaic2} show a number of the multi-wavelength signatures emitted by the 10 March 2001 solar flare. Frames \ref{mosaic2}a and \ref{mosaic2}b show the position of the magnetic transients, represented by the yellow and green circles, over the MDI-intensity continuum and magnetogram respectively. The magnetic neutral line is over-plotted (red line) in all frames for reference. Figure~\ref{mosaic2}c shows the magnetic difference maps at the time of the maximum of the flare (04:04:01.61~UT). We can clearly see one transient coincides well with the region of HXR emission (denoted by the contours), lying across the magnetic neutral lines. In Figure~\ref{mosaic2}d we have plotted the Doppler differences for the same time. Here we can see two photospheric signatures (spatially coinciding with the magnetic transients) that can be associated with surface perturbations of the solar photosphere. We also note that observations by \citet{lietal2005} show that the WL signature is composed of two sources, both of them being well correlated (spatially) with the magnetic transients. One strong and extended source lies in the region of the HXR and seismic source; the second one appears to correlate well with the second magnetic transient.

\citet{uddinetal2004} extensively analyzed the temporal and spatial behavior of the solar flare.  The maximum time in both HXR and MW emission reported by them as well as by \citet{lietal2005} (who undertook a very similar analysis), coincides very well with the maximum of the seismic emission (following the already well known delay of approximately three -- four minutes \citep{moradi07}). \citet{uddinetal2004} also discuss the spatial correlation between the different sources, showing that all three forms of emissions (WL, HXR and MW) are located in the region of maximum magnetic shearing. \citet{chandra06}, in a similar work, reported the locations of two $\mathrm H\alpha$ kernels in the flaring region. One of these kernels is (spatially) well correlated with the HXR source (observed by {\it Yohkoh}) and the observed seismic source, suggesting the precipitation of electrons in the chromosphere. The second $\mathrm H\alpha$ kernel is however not correlated with any HXR source, possibly indicating proton precipitation in this region (see \citet{zg2004} for a discussion about the partial separation of electrons and protons into the loop legs).


\section{The Magnetic Field}

The magnetic field topology of the active region has also been studied by other authors \citep{uddinetal2004, lietal2005} and was correlated with other emissions produced by the flare. Using vector magnetograms from the Mitaka Solar Observatory (Figure~\ref{vm}), we can see that the shearing of the magnetic field lines is close to $\mathrm 80^{\mathrm o}$ at the location of the seismic source (see the white arrow) - which would imply that a vast amount of energy was stored in the magnetic field prior to the flare. The area where the shearing is significant is very small. The seismic source itself is proved to be of a small size of $\mathrm 19 \times 25$~Mm. The magnetic energy released by the flare is used to accelerate particles, heat the chromosphere, and also drive the coronal mass ejection \citep[see][]{uddinetal2004,lietal2005}, and produce the compact seismic source.

\begin{figure}[!h]
\begin{center}
\includegraphics[width=0.75\columnwidth]{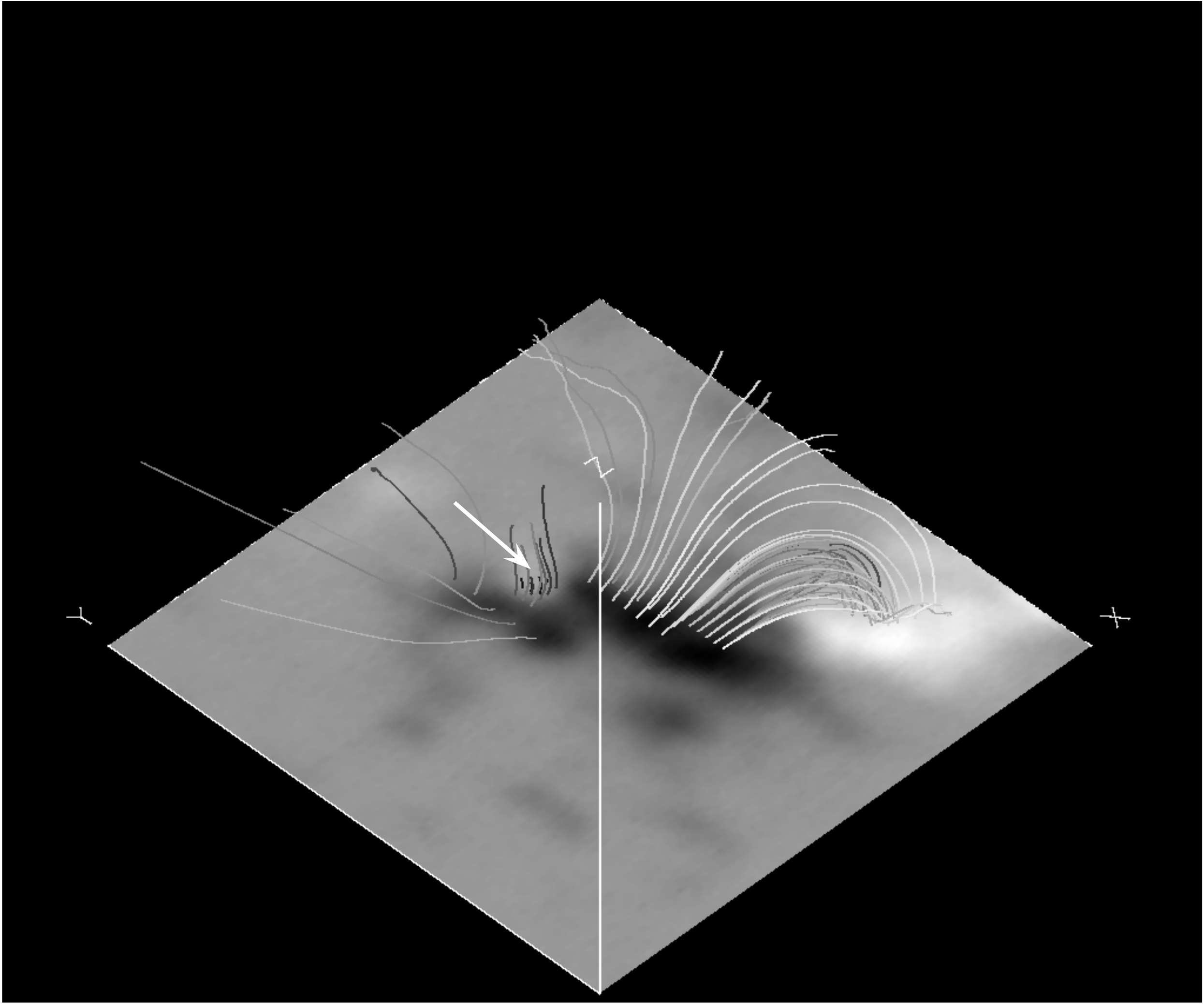}
\caption{NLFFF magnetic field extrapolation. The arrow shows the low-lying magnetic field region associated with the seismic emission.}
\label{extrapol}
\end{center}
\end{figure}

In order to verify the magnetic-field configuration of the active region (particularly in the corona), we computed the non-linear force free field (NLFFF) coronal magnetic field extrapolations of the active region using vector magnetograms from the Mitaka Solar Observatory. The resulting extrapolations (seen in Figure~\ref{extrapol}) clearly show high-altitude magnetic field lines connecting the two leading sunspots of the group, while between the leading and the following sunspots, only low-lying loops are visible (see arrow in Figure~\ref{extrapol}). A comparison between the extrapolations with SOHO/EIT images at 171\AA~ (Figure~\ref{mosaic2}f) and Figure~\ref{extrapol}) shows that our derived coronal magnetic-field extrapolations are in agreement with the observed magnetic field. Because of the close proximity of the sunspot to the solar limb and other observational constraints, it is not entirely possible to fully reconstruct the complete configuration of the magnetic field (in the flaring region). But nonetheless, we can qualitatively infer the overall structure of the coronal magnetic field from our estimates. 

In a closely related work, \citet{chandra06} conducted a detailed study of the dynamics of 10 March 2001 flare. As mentioned previously, they identified two $\mathrm H\alpha$ kernels, with only one kernel (K1) found to be spatially correlated with the HXR emission (see Figure~\ref{mosaic2}e) and therefore with the seismic source. The second $\mathrm H\alpha$ kernel, labelled K2, has an elongated structure. No HXR emission has been correlated with this source, we have also not detected any seismic source from this region despite the WL signature present at $\mathrm \approx$~04:04~UT.  These findings, along with observations of the flaring region made by the SXR telescope onboard {\it Yohkoh}, led the authors to propose a possible configuration of the magnetic field composed of two magnetic loops sharing one footpoint (``three-legged'' configuration), and associated with the single HXR source observed by {\it Yohkoh}. One of the loops appears to be connecting the shared footpoint with an opposite-polarity region associated by \citet{chandra06} with a secondary, stronger, yet distant MW source. The second loop is a low-lying loop connecting the shared footpoint with another located inside the region with a high degree of magnetic shearing. Furthermore, it is important to state that the two kernels observed by \citet{chandra06} spatially coincide remarkably well with the magnetic transients observed in Figure~\ref{mosaic2}e), with only one of them also being well correlated with the HXR emission and the Doppler signature.

The existence of a relationship between the height of the coronal magnetic loops and the seismicity of active regions has previously been proposed in \citet{Martinez-Oliveros2006}. The idea behind this assertion was that electrons in short, low-altitude magnetic loops precipitate more effectively than long, high-altitude loops because of enhanced scattering by thermal electrons ablated from the chromosphere.
Electrons whose pitch angles 
are greater than the loss-cone threshold, are trapped in the corona until they are scattered into the loss cone. Eventually, these electrons precipitate into the chromosphere and, depending on their energy, into the photosphere, transferring efficiently energy and momentum to the system. This scattering rate is greatly increased when the population of thermal electrons in the loop is large. This generally depends on the ablation of chromospheric gas into the corona by the fraction of electrons that were initially injected into the loss cone.  The volume of a short, low-lying loop is much smaller than that of a long high-altitude loop.  The electron density that results from a given mass of the chromosphere having been ablated is thus inversely proportional to this volume. Hence, given these understandings, we propose that short, low-lying loops become efficient scattering environments promptly greatly expediting precipitation on time scales conducive to seismic emission.

The collapse or relaxation of a high-altitude loop into a low-altitude one due to reconnection can greatly expand the loss cone, which would then enhance the precipitation distribution if pitch angles were left unchanged \citep{aschwanden2004}.
As we understand it, such a collapse facilitates electrons, initially trapped in the coronal magnetic field, to precipitate into the chromosphere and photosphere.
Observation in $\mathrm{H\alpha}$ \citep{uddinetal2004} of this flare, show the evolution of the filaments in the flaring region, changing from a potential configuration to a sigmoidal structure due of the high shearing of the magnetic field, with a post-flare relaxation of the magnetic field lines also observed in $\mathrm{H\alpha}$. This suggests that the above scenario of electron injection could very well take place, making the electron precipitation process much more efficient.

\section{Discussions}

The standard flare scenario divides the flare process into a number of phases. In this scenario, the flare particles are accelerated to relativistic or super-relativistic velocities in the corona and injected into magnetic field loops whose footpoints are in active-region chromospheres. Inevitably, some particles are going to be trapped in the coronal magnetic field, while others, those in the magnetic loss cone, will precipitate directly into the chromosphere. Eventually the majority of the trapped particles are either scattered into the loss cone and precipitated, or thermalized (or both) by thermal plasma in the magnetic loop. In the case of the very sudden and impulsive flare of 10 March 2001, the hypothesis is that acceleration and injection of particles into the magnetic loop occurred in a short period of time \citep{uddinetal2004}. 

This kind of phenomenon can be described using the trapping and injection model proposed by \citet{aschwanden2004}. In this model the rate of precipitation of charged particles into the chromosphere is controlled by the relaxation time of the system. The aperture angle of the loss cone changes with time, significantly opening as the magnetic field collapses to a more potential configuration. It is important to note that in this model, the time of acceleration and injection of the particles into the magnetic field are almost the same and relatively short compared with the precipitation and trapping time. It is fair to assume that if the relaxation time is short, the aperture of the loss cone also will change rapidly, allowing more particles to reach the chromosphere in a short period of time. This depends on efficient scattering of high-energy electrons into the expanded loss cone, which is greatly enhanced by chromospheric ablation of thermal plasma into short, low-lying loops.  Rapid evacuation of trapped electrons is suggested by observations of a rapid decay in non-thermal MW emission.  As a general rule, thermalization of particles in a magnetic trap is small compared to losses due to precipitation. Hence, high-energy electrons evacuated from the coronal loop in this way contribute to HXR bremsstrahlung emission substantially as well as their counterparts that were initially injected into the loss cone.


A much more complex model of particle precipitation, in which processes such as non-thermal excitation and ionization of hydrogen atoms, and non-thermal plasma heating (coulomb and ohmic) is explored by \citet{ah1986}, \citet{zk1993}, and \citet{zharkova2007}. Interestingly, the later show the ohmic heating of the corona by the electron beams is so effective that the corresponding particle-induced downward propagating shocks are almost depleted of energy, leaving very little energy to reach the photosphere and induce any kind of seismic activity. Perhaps this is the explanation for why we did not see any seismic sources at the location of the $\mathrm H\alpha$ K2 kernel in Figure~\ref{egression}. However, we also want emphasize here the possibility that photospheric heating also contributes to flare acoustic emission.

As the whole flaring process occurred relatively rapidly (and given the highly-impulsive properties of the 10 March 2001 flare, it is not unreasonable to assume so), the solar chromosphere was heated quite suddenly. As we have already seen, the multi-wavelength emissions of the flare indicate this; furthermore the strong spatial and temporal correspondence between the different types of emissions point to radiative back-warming playing a significant role in the heating mechanism. This conclusion was in fact drawn by both \citet{lietal2005} and \citet{ding03} to explain the origin of the continuum feature of the 10 March 2001 flare in terms of an ``electron-beam-heated flare model'', with chromospheric radiative back-warming suspected being the chief heating agent, originating in the temperature-minimum region.

The above conclusions, when viewed in conjunction with those of \citet{donea05}, \citet{donea06}, \citet{moradi07}, and \citet{Martinez-Oliveros2006}, provide direct evidence of flare acoustic emission being driven, in part, by heating of the low photosphere. The basic principle here is that the chromospheric radiation further heats up the photosphere, with the result being of an optically-thick $\mathrm H^{-}$ bound-free absorption, which then introduces a pressure transient directly to the underlying medium. The photospheric heating hypothesis is well supported by our observations and previous ones -- which all indicate that instances of flare seismic emission have been characterized by a close spatial correspondence between the seismic emission and sudden WL emission during the impulsive phase of the flare. Radiative fluxes characteristic of WL emission seen in all acoustically active flares, if emitted downward from the chromosphere (as well as upward), are probably sufficient to heat the photosphere a few percent within a few seconds of the onset of the incoming radiative flux. This process, described by \citet{machado1989}, is called ``back-warming''. 

According to rough models described by \citet{donea06} and \citet{moradi07}, such heating (if applied suddenly) should cause a pressure transient in the heated layer that drives a seismic transient whose energy flux is of the order of those estimated for acoustically active flares. The energy invested into the seismic transient is in proportion to
\begin{equation}
\varepsilon \sim (\frac{\Delta I_c}{I_c})^2
\end{equation}
a fraction of order $\Delta I_c/I_c$ times the radiative energy suddenly emitted by the flare. We therefore expect acoustic emission due to photospheric heating to be inefficient in flares whose WL signatures are weak, diffuse, or not very sudden, and this is consistent with examples we have encountered to date.

In conclusion, it must be stated that to date, no one, single mechanism can fully explain the mechanics of flare acoustics and their observational signatures - because to do so would be a gross over-simplification of the problem. What these results have shown is that the study of flare mechanics, as well as helioseismology, would greatly benefit from the development of detailed models of solar flare-induced seismic emission -- from the corona, down to the photosphere, including modeling of realistic active region subphotospheres. Credible models would need to include realistic subphotospheric thermal anomalies to represent penumbral and perhaps umbral subphotospheres, and realistic photospheric magnetic fields extrapolated to depths of a few hundred kilometers beneath the photosphere. The later should also include an account for the highly inclined magnetic fields that characterize sunspot penumbra -- where a significant majority of the seismic emissions observed to date have been detected.

\thanks{The authors would like to sincerely thank Profs. Paul Cally, Markus Aschwanden, Valentina Zharkova, Drs. Charlie Lindsey and Wahab Uddin, and Diana Besliu-Ionescu for their helpful and interesting comments which contributed directly to the development and/or improvement of this article.}



\begin{thebibliography}{}

\bibitem[\protect\citeauthoryear{Aboudarham and Henoux}{1986}]{ah1986} Aboudarham, J., Henoux, J.C.: 1986, \aap~{\bf 186}, 73

\bibitem[\protect\citeauthoryear{Aschwanden}{2004}]{aschwanden2004} Aschwanden M.: 2004, \apj~{\bf 608}, 554

\bibitem[\protect\citeauthoryear{Besliu-Ionescu \etal}{2006}]{betal2006}
Besliu-Ionescu, D., Donea, A.-C., Cally, P. S. and Lindsey, C.: 2006, In: a new era in helio-and asteroseismology, Proceedings of the 2006 SOHO-18/GONG-2006/HELAS I Meeting, Fleck B. (ed.), ESA Publications, Darmstadt, CDROM, 67.1.

\bibitem[\protect\citeauthoryear{Chandra \etal}{2006}]{chandra06} Chandra, R., Jain, R., Uddin, W., Yoshimura, K., Kosugi, T., Sakao, T., Joshi, A., Deshpande, M.R.: 2006, \solphys~{\bf 239}, 239

\bibitem[\protect\citeauthoryear{Ding \etal}{2003}]{ding03} Ding, M. D.,Liu, Y., Yeh, C.-T., Li, J. P.: 2003, \aap~{\bf 403}, 1151

\bibitem[\protect\citeauthoryear{Donea and Lindsey}{2005}]{donea05} Donea, A.-C., Lindsey, C.: 2005 \apj~{\bf 630}, 1168

\bibitem[\protect\citeauthoryear{Donea, Braun and Lindsey}{1999}]{donea99} Donea, A.-C., Braun, D.C, Lindsey, C.: 1999, \apj~{\bf 513}, L143

\bibitem[\protect\citeauthoryear{Donea \etal}{2006}]{donea06} Donea, A.-C., Be\c{s}liu-Ionescu, 
D., Lindsey, C. \& Zharkova, V.V.: 2006, \solphys~{\bf 239}, 113

\bibitem[\protect\citeauthoryear{Kosovichev and Zharkova}{1998}]{kz1998} Kosovichev, A., Zharkova, V.V.: 1998 \nat~{\bf 393}, 317

\bibitem[\protect\citeauthoryear{Li, Ding and Liu}{2005}]{lietal2005} Li, J.P., Ding D., Liu, Y.: 2005, \solphys~{\bf 229}, 115

\bibitem[\protect\citeauthoryear{Lindsey and Braun}{2000}]{lb2000}
Lindsey, C., Braun, D.C.: 2000, \solphys~{\bf 192}, 261

\bibitem[\protect\citeauthoryear{Liu, Ding and Fang}{2001}]{liuetal2001} Liu, Y., Ding, M.D., Fang, C: 2001, \apj~{\bf 563}, L169

\bibitem[\protect\citeauthoryear{Machado, Emslie and Avrett}{1989}]{machado1989}
Machado, M.E., Emslie, A.G. and Avrett, E.H.: 1989, \solphys~{\bf 124} 303

\bibitem[\protect\citeauthoryear{Mart\'inez-Oliveros \etal}{2007}]{Martinez-Oliveros2006}  Mart\'inez-Oliveros, J.C., Moradi, H., Besliu-Ionescu, D.,Donea A.-C, Cally, P.S.: 2007, \solphys~{\bf 245}, 121

\bibitem[\protect\citeauthoryear{Moradi \etal}{2007}]{moradi07} Moradi, H., Donea, A.-C., Lindsey, C., Be\c{s}liu-Ionescu, D., Cally, P.S.: 2007, \mnras~{\bf 374}, 1155


\bibitem[\protect\citeauthoryear{Rajaguru \etal}{2006}]{raj2006} Rajaguru, S.P., Birch, A.C., Duvall, T.L., Jr., Thompson, M.J., Zhao, J.: 2006, \apj~{\bf 646}, 543

\bibitem[\protect\citeauthoryear{Uddin \etal}{2004}]{uddinetal2004} Uddin, W., Jain, R., Yoshimura, K., Chandra, R., Sakao, T., Kosugi, T., Joshi, A., Despande, M.R.: 2004, \solphys~{\bf 225}, 325

\bibitem[\protect\citeauthoryear{Zharkova and Gordovskyy}{2004}]{zg2004} Zharkova, V.V., Gordovskyy, M..: 2004, \apj~{\bf 604}, 884

\bibitem[\protect\citeauthoryear{Zharkova and Kobylinskii}{1993}]{zk1993} Zharkova, V.V., Kobylinskii, V.A..: 1993, \solphys~{\bf 143}, 259

\bibitem[\protect\citeauthoryear{Zharkova and Zharkov}{2007}]{zharkova2007} Zharkova V.V., Zharkov S.: 2007, \apj~{\bf 664}, 573

\end{thebibliography}
\end{document}